# Giant spin Hall effect in AB-stacked MoTe$_2$/WSe$_2$ bilayers

Zui Tao[1*], Bowen Shen[1*], Wenjin Zhao[2*], Nai Chao Hu[3], Tingxin Li[1], Shengwei Jiang[1], Lizhong Li[1], Kenji Watanabe[4], Takashi Taniguchi[4], Allan H. MacDonald[3], Jie Shan[1,2,5], Kin Fai Mak[1,2,5]

[1]School of Applied and Engineering Physics, Cornell University, Ithaca, NY, USA
[2]Kavli Institute at Cornell for Nanoscale Science, Ithaca, NY, USA
[3]Department of Physics, University of Texas at Austin, Austin, TX, USA
[4]National Institute for Materials Science, 1-1 Namiki, 305-0044 Tsukuba, Japan
[5]Laboratory of Atomic and Solid-State Physics, Cornell University, Ithaca, NY, USA

*These authors contributed equally
Email: jie.shan@cornell.edu; kinfai.mak@cornell.edu

**The spin Hall effect (SHE), in which electrical current generates transverse spin current, plays an important role in spintronics for the generation and manipulation of spin-polarized electrons [1–7]. The phenomenon originates from spin-orbit coupling. In general, stronger spin-orbit coupling favors larger SHEs but shorter spin relaxation times and diffusion lengths [1,4–7]. To achieve both large SHEs and long-range spin transport in a single material has remained a challenge. Here we demonstrate a giant intrinsic SHE in AB-stacked MoTe$_2$/WSe$_2$ moiré bilayers by direct magneto optical imaging. Under moderate electrical currents with density < 1 A/m, we observe spin accumulation on transverse sample edges that nearly saturates the spin density. We also demonstrate long-range spin Hall transport and efficient non-local spin accumulation limited only by the device size (about 10 μm). The gate dependence shows that the giant SHE occurs only near the Chern insulating state, and at low temperatures, it emerges after the quantum anomalous Hall breakdown. Our results demonstrate moiré engineering of Berry curvature and large SHEs for potential spintronics applications.**

**Main**
The spin Hall effect has two distinct mechanisms, intrinsic and extrinsic, in a material [7]. The intrinsic effect is caused by spin-orbit coupling in the material's electronic band structure, whereas the extrinsic effect is caused by spin-orbit coupling between electrons and impurities through scattering. Materials with large intrinsic SHEs are sought after for spintronics applications [4–7]. The intrinsic spin Hall conductivity is directly related to the spin-dependent Berry curvature, which acts like a magnetic field in momentum space. Particularly, monolayer transition metal dichalcogenides (TMDs) with strong Ising spin-orbit coupling exhibit opposite Berry curvature at the K and K' valleys, which is also spin dependent due to spin-momentum locking [8–12]. The non-zero Berry curvature has manifested a valley and spin Hall effect in doped monolayer TMD semiconductors [13,14]. But the effect is weak because Berry curvature spreads over large momentum space near the K (K') point in the case of large band gaps. Tailoring the electronic band structure by forming moiré heterostructures, specifically creating gapped band crossings with small gaps near the Fermi level, opens a new route to generate Berry curvature hotspots and large SHEs [15–20].

Moiré materials are a highly tunable electronic system, in which rich quantum phenomena from the effects of strong electronic correlations and nontrivial band topology have been realized [16–20]. One such example is AB-stacked (60-degree-aligned) $MoTe_2/WSe_2$ moiré bilayers. The TMD bilayers form a triangular moiré lattice. But the Wannier orbitals of the topmost $MoTe_2$ and $WSe_2$ moiré valence states are centered at two different sublattice sites and form an effective honeycomb lattice [21–23]. The low-energy physics can be mapped to the Kane-Mele model [24,25] with interlayer (nearest-neighbor) hopping, complex intralayer (next-neighbor) hopping, and a sublattice potential difference that is continuously tunable by an out-of-plane electric field [21–23,26–29]. The electric field can induce moiré band inversion and topological bands [30,31]. Recent experiments have shown that at filling factor $\nu = 1$ of the moiré unit cell, electron-electron interactions can further induce ferromagnetic instability and a Chern insulator [31] with spin degeneracy spontaneously lifted in each TMD layer [32]. In addition, long spin relaxation time (> 8 μs) near $\nu = 1$, presumably also from the strong electron-electron interactions, has been observed in related TMD moiré bilayers [33] (independent of the Chern state). These results suggest a unique opportunity to realize both strong SHEs and long-range spin transport in moiré bilayers.

**The spin Hall effect**
We fabricate dual-gated field-effect devices of AB-stacked $MoTe_2/WSe_2$ bilayers. Figure 1b shows an optical image of a typical device. The device channel is shaped into Hall bar geometry for simultaneous electrical transport and magneto optical imaging studies (Fig. 1a). The channel has typical width and length of 3 and 9 μm, respectively, and is defined by the top gate and the contact electrodes. The top and bottom gate voltages ($V_{tg}$ and $V_{bg}$) independently control the hole filling factor $\nu$ in the moiré bilayer and the out-of-plane electric field $E$ across it. To detect the SHE, we employ magnetic circular dichroism (MCD) imaging at 730 nm, which is on resonance with the fundamental intralayer exciton of the $WSe_2$ layer. The MCD is proportional to the spin and valley polarization of the $WSe_2$ layer and is used to characterize the spin density $n_s$ in the bilayer [34,35]. More details on device fabrication and measurements are provided in Methods.

Figure 1c shows the four-terminal longitudinal resistance $R_{xx}$ as a function of the two gate voltages at temperature $T = 1.6$ K. The arrows label the electric field and filling factor axes. At fixed filling factor $\nu = 1$ (black dashed line), for small electric fields, the system is a charge-transfer insulator with divergent $R_{xx}$ at low temperatures. For intermediate electric fields, the system is a Chern insulator with small $R_{xx}$ (denoted by a dashed circle). It supports an intrinsic quantum anomalous Hall (QAH) effect (Extended Data Fig. 1), where the Hall resistance $R_{xy}$ is quantized and the longitudinal resistance vanishes at small magnetic fields. Both quantities exhibit magnetic hysteresis. The MCD is correlated with the transport characteristics, and also exhibits a spontaneous signal and magnetic hysteresis. These results are consistent with earlier studies [31].

Figure 1d shows the MCD image of the device at 1.6 K, zero magnetic field and zero bias. The gate voltages are parked in the middle of the Chern insulator state. Positive MCD (or negative MCD if the spins are polarized in the opposite direction) is observed throughout the channel. The spatial variations in the MCD magnitude reflect inhomogeneous magnetization in the bilayer [15], which may arise from strain and/or twist angle disorders [36] (see Extended Data Fig. 2 for further characterization of the sample inhomogeneity). At 6 K, MCD vanishes (see Extended Data Fig. 3 for images at intermediate temperatures). This is consistent with the reported ferromagnetic ordering temperature $T_c \approx 5$ K of the Chern insulator state [31,32]. However, when the device is biased at 3.3 µA along the long axis, we observe MCD of opposite signs on two edges of the channel and zero MCD in the middle of the channel. MCD on the two edges switches sign when the bias current is reversed (Extended Data Fig. 4).

The observed effect above $T_c$ is consistent with the SHE (Fig. 1a). The electrical current with density $J$ along the long axis drives a transverse spin current with density $J_s$ along the short axis of the Hall bar, which pumps opposite spins onto the two edges. Under steady-state conditions, spin pumping is balanced by spin relaxations. This results in spin accumulation on the edges with a characteristic decay length $l_s$ into the bulk that is governed by spin diffusion [2,37]. In Fig. 1d, however, we do not observe any clear decay of the MCD from the edges. The channel is nearly divided into two large domains of opposite spins. This suggests that spin diffusion is limited by the channel width (see below).

Below we fix the gate voltages in the middle of the Chern state and characterize the SHE above $T_c$ and the interplay between the SHE and QAH below $T_c$. We then study the gate dependence of the SHE to explore its origin. In the latter, we employ measurement geometry shown in the lower panel of Fig. 1a to increase the device length for long-range spin Hall transport. Here the electrical current is biased along the short axis and the spin current is driven along the long axis of the Hall bar. Unless otherwise specified, the magnetic field is fixed at zero.

**Interplay between the SHE and QAH**
Figure 2a shows the MCD images at representative currents biased along the long axis of the device at 6 K (see Extended Data Fig. 5 for images with currents biased along the short axis). The MCD amplitude increases monotonically with the current. Figure 2c (top panel) summarizes the current dependence of MCD at two edges, $P_1$ and $P_2$, as denoted in Fig. 2a. The response is linear for small currents and saturates above about 1 µA (corresponding to a density of 0.3 A/m). The saturation MCD is comparable to that measured at zero bias under high magnetic fields (Extended Data Fig. 6), where the gate-doped holes are almost fully spin-polarized. The observed SHE is thus highly efficient; with moderate currents it nearly saturates the spin density in the sample. In comparison, the SHE-induced edge magnetization reported in conventional semiconductors is orders of magnitude smaller than the saturation value [2,3,13,14,37]. The SHE in moiré bilayers also has weak temperature dependence (Fig. 2d). The study is performed up to 18 K because the exciton resonance that is used to probe the MCD remains temperature independent.

Next, we study the bias dependent MCD images at 1.6 K, at which the Chern insulator emerges (Fig. 2b). As the current increases, a small domain of negative MCD first appears near the top right corner of the image and grows gradually in size. Eventually, the sample is split into two large domains of opposite magnetizations on each side of the current (Supplementary MCD movies). The resulting MCD image is nearly identical to that at 6 K under high bias, and is independent of the spontaneous magnetization direction (Extended Data Fig. 4). The high-bias MCD at 1.6 K is consistent with the SHE-induced spin accumulation discussed above.

Figure 2c summarizes the bias dependence of MCD at $P_1$ and $P_2$ and the device transport characteristics at 1.6 K. When the current reaches about 0.5 $\mu A$, the MCD at one of the edges ($P_1$) changes sign. Concurrently, the spontaneous Hall response diminishes and the longitudinal resistance increases substantially, indicating a QAH breakdown. The QAH breakdown corresponds to the termination of edge-dominant transport and the onset of bulk transport. The SHE, manifested in spin accumulation and MCD sign change at $P_1$, becomes relevant only beyond the QAH breakdown.

The bias dependent MCD at 1.6 K consists of two contributions, the spontaneous magnetization of the Chern insulator and the spin accumulation from the SHE. In Fig. 2e, we extract the spontaneous magnetization contribution as the difference between MCD at 1.6 K and 6 K because MCD at 6 K is purely from the SHE, which is weakly temperature dependent. This analysis is further justified by the nearly identical high-bias MCD images at 1.6 K and 6 K (Fig. 2a,b and Supplementary MCD movies). The spontaneous magnetization is quenched above ~ 0.5 $\mu A$, which agrees with the QAH breakdown observed in transport studies (Fig. 2c). The relevance of QAH breakdowns is further supported by the magnetic field dependent breakdown current (Extended Data Fig. 7). A higher current is required for QAH breakdown since the out-of-plane magnetic field helps to stabilize the QAH effect.

**Gate dependence**
We examine the gate dependence of the SHE to further elucidate its relationship with the Chern insulating state. We scan $V_{bg}$ with $V_{tg}$ fixed at -4.692V. Figure 3a shows the MCD images at representative filling factors at 1.6 K. The bias current is 2.6 µA so that the system is far beyond the QAH breakdown (Extended Data Fig. 5). We bias the device along the short axis near one end of the Hall bar to increase the spin transport length. The current is confined to a small region near the tip of the electrodes. Opposite magnetizations are observed on two sides of the current. The current path, identified from zero MCD, deviates from a straight line due to sample inhomogeneities. (Extended Data Fig. 8 shows more MCD images for current biased with different electrode pairs.) Overall, the current-induced magnetization is the strongest at $\nu = 1$ and drops sharply both above and below $\nu = 1$. At the same time, the induced magnetization spreads the furthest from the source at $\nu = 1$.

We analyze the MCD, or equivalently, the spin density $n_s$ to the right of the current. Figure 3b illustrates a line profile of MCD along the grey line in Fig. 3a. The spin density arises from the source (vertical dashed lines), peaks in the middle, and vanishes at the end of the channel because of the presence of an end electrode (Fig. 1b). A similar behavior is observed for the spin density to the left of the source. We illustrate the filling dependence of the spin density at two representative locations, P_3 and P_4, in Fig. 3d. Spin accumulation is sharply peaked around $\nu = 1$ at P_4 which is 6 μm away from the source. In contrast, closer to the source at P_3, the distribution is wider; in addition to the peak around $\nu = 1$, a second weaker peak is also observed around $\nu = 1.2$.

We extract the steady-state spin current density, $J_s$, from the measured spin density by using the continuity equation, $\frac{\partial J_s}{\partial x} + \frac{n_s}{\tau_s} = 0$. Here $x$ is the propagation distance of the spin current, and $\tau_s$ is the spin relaxation time. We numerically integrate the MCD line profile with the boundary condition that $J_s$ vanishes at the right end of the device. The spin current density is normalized to 1 at the source (Fig. 3c). It decays towards the right end of the device. Deviations from smooth monotonic decay are likely effects of sample inhomogeneity.

Figure 3e illustrates the filling dependence of the spin Hall conductivity $\sigma_{SH}$ (that is, the spin current density per unit bias electric field). We characterize $\sigma_{SH}$ using $J_s$ at the source normalized by the sample resistance because in the measurement the current is kept constant for different filling factors, and the bias electric field is proportional to the sample resistance. The spin Hall conductivity is peaked around $\nu = 1$ with a second weaker peak around $\nu = 1.2$. We also use the width at half-maximum of the $J_s$ line profile to characterize the spin transport distance (Fig. 3f). The width increases sharply with filling factor, peaks around $\nu = 1$, and gradually decreases for $\nu > 1$. The maximum width (~ 5 μm) is comparable to the channel length (~ 9 μm) and is therefore an underestimate of the spin diffusion length $l_s$ (see Methods for modeling of spin diffusion in a rectangular slab and Extended Data Fig. 9). Future studies using a longer and more homogeneous device channel are required to determine $l_s$ near $\nu = 1$.

**Discussions and conclusions**
We have observed a giant SHE in AB-stacked MoTe_2/WSe_2 moiré bilayers by direct MCD imaging. In contrast to conventional semiconductors, the SHE-induced magnetization in TMD moiré bilayers can nearly saturate the spin density for moderate bias currents. The effect is dominated by the intrinsic Berry curvature contribution. This is supported by both the weak temperature dependence of the effect (Fig. 2d) and the significantly enhanced $\sigma_{SH}$ at $\nu = 1$, where a Chern insulator emerges. The Berry curvature hotspots are expected from interaction-driven gap opening at $\nu = 1$ (Ref. [21,31]). Secondary Berry curvature hotspots are also possible, which could explain the observed smaller enhancement of $\sigma_{SH}$ near $\nu = 1.2$. But further investigations are required to verify it. The large SHE-induced magnetization at $\nu = 1$ in the steady state is also enabled by the long spin relaxation time (and spin diffusion length). We have observed spin diffusion over several μm's, which is limited by the device channel length. Our results highlight the potential of moiré engineering for creating Berry curvature hotspots. The demonstration of giant SHEs in the same material platform as many reported exotic quantum many-body phenomena opens exciting opportunities for gate-defined lateral heterostructure quantum devices.

**Methods**
**Device fabrication**
We fabricated AB-stacked MoTe$_2$/WSe$_2$ Hall bar devices using the layer-by-layer dry transfer method [38]. In short, we first exfoliated the bulk crystals onto Si substrates to obtain MoTe$_2$ and WSe$_2$ monolayers and few-layer hexagonal boron nitride (hBN) and graphite. The crystal orientations of the MoTe$_2$ and WSe$_2$ monolayers were determined by angle-resolved optical second-harmonic generation spectroscopy before stacking. The constitute layers were then picked up sequentially using a polycarbonate (PC) stamp to realize the dual-gated device, in which the TMD bilayers are encapsulated between hBN and graphite top and bottom gates [31]. We used a relatively thin (approximately 5 nm) hBN layer for the top gate to support a large out-of-plane electric field. We used 5-nm-thick Pt as metal electrodes for better electrical contacts to the sample while keeping the strain effects minimal. Two different devices were examined in this study; the results are reproducible.

**Electrical transport measurements**
Both the electrical transport measurements and optical imaging were performed in a closed-cycle $^4$He cryostat (attoDRY 2100) equipped with a 9 T superconducting magnet. The base temperature is about 1.6 K. We applied the low-frequency (17.717 Hz) lock-in techniques to measure the sample resistances. For zero DC bias, a small AC bias of 0.7 mV was applied to obtain the longitudinal resistance in Fig. 1c; the resultant AC current was kept under 20 nA to avoid heating and high bias effects. For transport under a DC bias, a function generator and dividers (100:2 and 100:3) were used to superpose a small AC current on the DC current. The AC current was kept below 5 nA. After a low-noise current preamplifier (Stanford Research SR570), the DC component of the current was measured by a multimeter (Keithley 2000) and the AC component was measured by a lock-in amplifier to yield the differential resistances.

**MCD imaging**
A light-emitting diode (LED, Thorlabs M730L5) with nominal wavelength at 730 nm was used as the illumination source. Two tunable bandpass filters were used to further engineer the light spectrum to be centered at the A exciton resonance of WSe$_2$ with 3-nm full-width-half-maximum. A Glan-Taylor polarizer and a broadband quarter-wave plate were combined to change the light helicity between the left and right circular polarization. A low-temperature objective of numerical aperture 0.8 was used to focus light on the sample. The incident light intensity on sample was kept below 1 nW/μm$^2$ to minimize the heating effect on the sample. The reflected light was collected by the same objective and sent to a 2D electron-multiplying charge-coupled device (Princeton Instruments, ProEM 512 × 512) for imaging. The typical integration time for an image was 0.5 s. The MCD $\left(\frac{I^- - I^+}{I^- + I^+}\right)$ was determined as the relative intensity difference between the image of left ($I^+$) and right ($I^-$) circularly polarized light. A background MCD signal of about 0.3% was present in the setup. It arises from polarization imperfections of the optics such as the beam splitter and objective. We used the MCD image of the sample in the heavily hole-doped regime, in which the sample spectrum is featureless, as the background and subtracted it from all MCD images.

**Spin continuity equation and spin current**
In the moiré bilayers investigated here, the spin diffusion model is valid (the spin relaxation time far exceeds the momentum relaxation time). We model our device as a rectangular slab with width $W$ and length $L$ as shown in Extended Data Fig. 9a. The electrical current is applied along the y-direction in the middle of the device; the dark blue dots mark the current ejection/extraction points. The spin current is driven along the x-direction. The origin of the coordinates is set at the center of the slab. Following Ref. [39], we solve the damped diffusion equation with the appropriate boundary conditions to obtain an analytical expression for the steady-state spin density $n_s(x,y)$:

$$n_s(x,y) \sim \int_{-\infty}^{+\infty} \frac{dk}{2\pi} e^{ikx} \frac{\cosh[\omega(k)y]}{\omega(k)\coth[kW/2]\sinh[\omega(k)W/2] + k\tan^2\theta_H \cosh[\omega(k)W/2]}.$$

The solution is expressed in terms of inverse Fourier transforms. Here parameter $\omega(k) = \sqrt{k^2 + l_s^{-2}}$ depends on the spin diffusion length $l_s$; $\theta_H$ denotes the Hall angle and $\tan(\theta_H)$ is the ratio of the spin Hall conductivity and the ordinary longitudinal charge conductivity. The spin current density $J_s$ can be obtained from the spin density profile using the spin continuity equation $\frac{\partial J_s}{\partial x} + \frac{\partial n_s}{\partial t} = -\frac{n_s}{\tau_s}$. In the steady state, the spin pumping rate balances the spin decay rate so that $\frac{\partial n_s}{\partial t} = 0$. The spin continuity equation becomes $\frac{\partial J_s}{\partial x} + \frac{n_s}{\tau_s} = 0$. The spin current density $J_s$ can therefore be obtained by integrating $n_s(x)$ at a fixed y and requiring that $J_s$ vanishes at the boundary of the channel.

We plot the spin density profile at $y = 0$ for different sample geometries and spin diffusion lengths in Extended Data Fig. 9b-d. A Hall angle $\theta_H = \pi/6$ is used. With increasing sample length $L$, the spin density $n_s$ appears more localized near the source and the spin current $J_s$ becomes more distinguishable for different diffusion lengths. The sample geometry in the experiment is closest to the case of $W = 3\ \mu m, L = 9\ \mu m$ in Extended Data Fig. 9c, which qualitatively reproduces the experimental results in Fig. 3.

**References**

1. Jungwirth, T., Wunderlich, J. & Olejník, K. Spin Hall effect devices. *Nat. Mater.* **11**, 382–390 (2012).
2. Kato, Y. K., Myers, R. C., Gossard, A. C. & Awschalom, D. D. Observation of the Spin Hall Effect in Semiconductors. *Science* **306**, 1910–1913 (2004).
3. Wunderlich, J., Kaestner, B., Sinova, J. & Jungwirth, T. Experimental Observation of the Spin-Hall Effect in a Two-Dimensional Spin-Orbit Coupled Semiconductor System. *Phys. Rev. Lett.* **94**, 047204 (2005).
4. Liu, L. *et al.* Spin-Torque Switching with the Giant Spin Hall Effect of Tantalum. *Science* **336**, 555–558 (2012).
5. Liu, L., Lee, O. J., Gudmundsen, T. J., Ralph, D. C. & Buhrman, R. A. Current-Induced Switching of Perpendicularly Magnetized Magnetic Layers Using Spin Torque from the Spin Hall Effect. *Phys. Rev. Lett.* **109**, 096602 (2012).
6. Zutic, I., Fabian, J., & Sarma, S. D. Spintronics: Fundamentals and applications. *Rev Mod Phys* **76**, 88 (2004).
7. Sinova, J., Valenzuela, S. O., Wunderlich, J., Back, C. H. & Jungwirth, T. Spin Hall effects. *Rev. Mod. Phys.* **87**, 1213–1260 (2015).
8. Xu, X., Yao, W., Xiao, D. & Heinz, T. F. Spin and pseudospins in layered transition metal dichalcogenides. *Nat. Phys.* **10**, 343–350 (2014).
9. Xiao, D., Yao, W. & Niu, Q. Valley-Contrasting Physics in Graphene: Magnetic Moment and Topological Transport. *Phys. Rev. Lett.* **99**, 236809 (2007).
10. Yao, W., Xiao, D. & Niu, Q. Valley-dependent optoelectronics from inversion symmetry breaking. *Phys. Rev. B* **77**, 235406 (2008).
11. Xiao, D., Liu, G.-B., Feng, W., Xu, X. & Yao, W. Coupled Spin and Valley Physics in Monolayers of MoS 2 and Other Group-VI Dichalcogenides. *Phys. Rev. Lett.* **108**, 196802 (2012).
12. Mak, K. F., Xiao, D. & Shan, J. Light–valley interactions in 2D semiconductors. *Nat. Photonics* **12**, 451–460 (2018).
13. Lee, J., Wang, Z., Xie, H., Mak, K. F. & Shan, J. Valley magnetoelectricity in single-layer MoS2. *Nat. Mater.* **16**, 887–891 (2017).
14. Mak, K. F., McGill, K. L., Park, J. & McEuen, P. L. The valley Hall effect in MoS2 transistors. *Science* **344**, 1489–1492 (2014).
15. Tschirhart, C. L. *et al.* Intrinsic spin Hall torque in a moire Chern magnet. (2022) doi:10.48550/arXiv.2205.02823.
16. Andrei, E. Y. *et al.* The marvels of moiré materials. *Nat. Rev. Mater.* **6**, 201–206 (2021).

17. Andrei, E. Y. & MacDonald, A. H. Graphene bilayers with a twist. *Nat. Mater.* **19**, 1265–1275 (2020).
18. Liu, J. & Dai, X. Orbital magnetic states in moiré graphene systems. *Nat. Rev. Phys.* **3**, 367–382 (2021).
19. Mak, K. F. & Shan, J. Semiconductor moiré materials. *Nat. Nanotechnol.* **17**, 686–695 (2022).
20. Kennes, D. M. *et al.* Moiré heterostructures as a condensed-matter quantum simulator. *Nat. Phys.* **17**, 155–163 (2021).
21. Devakul, T. & Fu, L. Quantum Anomalous Hall Effect from Inverted Charge Transfer Gap. *Phys. Rev. X* **12**, 021031 (2022).
22. Yang Zhang, Devakul, T. & Fu, L. Spin-textured Chern bands in AB-stacked transition metal dichalcogenide bilayers. *Proc. Natl. Acad. Sci.* **118**, e2112673118 (2021).
23. Rademaker, L. Spin-Orbit Coupling in Transition Metal Dichalcogenide Heterobilayer Flat Bands. *Phys. Rev. B* **105**, 195428 (2022).
24. Kane, C. L. & Mele, E. J. $Z_2$ Topological Order and the Quantum Spin Hall Effect. *Phys. Rev. Lett.* **95**, 146802 (2005).
25. Kane, C. L. & Mele, E. J. Quantum Spin Hall Effect in Graphene. *Phys. Rev. Lett.* **95**, 226801 (2005).
26. Fengcheng Wu, Lovorn, T., Tutuc, E., Martin, I. & MacDonald, A. H. Topological Insulators in Twisted Transition Metal Dichalcogenide Homobilayers. *Phys. Rev. Lett.* **122**, 086402 (2019).
27. Pan, H., Xie, M., Wu, F. & Das Sarma, S. Topological Phases in AB-Stacked $MoTe_2/WSe_2$: $Z_2$ Topological Insulators, Chern Insulators, and Topological Charge Density Waves. *Phys. Rev. Lett.* **129**, 056804 (2022).
28. Xie, Y.-M., Zhang, C.-P., Hu, J.-X., Mak, K. F. & Law, K. T. Valley-Polarized Quantum Anomalous Hall State in Moiré $MoTe_2 / WSe_2$ Heterobilayers. *Phys. Rev. Lett.* **128**, 026402 (2022).
29. Mai, P., Zhao, J., Feldman, B. E. & Phillips, P. W. 1/4 is the new 1/2: Interaction-induced Quantum Anomalous and Spin Hall Mott Insulators. Preprint at https://doi.org/10.48550/arXiv.2210.11486 (2022).
30. Zhao, W. *et al.* Realization of the Haldane Chern insulator in a moiré lattice. Preprint at http://arxiv.org/abs/2207.02312 (2022).
31. Tingxin Li *et al.* Quantum anomalous Hall effect from intertwined moiré bands. *Nature* **600**, 641–646 (2021).
32. Tao, Z. *et al.* Valley-coherent quantum anomalous Hall state in AB-stacked $MoTe_2/WSe_2$ bilayers. Preprint at https://doi.org/10.48550/arXiv.2208.07452 (2022).
33. Regan, E. C. *et al.* Mott and generalized Wigner crystal states in $WSe_2/WS_2$ moiré superlattices. *Nature* **579**, 359–363 (2020).
34. Tang, Y. *et al.* Simulation of Hubbard model physics in $WSe_2/WS_2$ moiré superlattices. *Nature* **579**, 353–358 (2020).

35. Tingxin Li *et al.* Continuous Mott transition in semiconductor moiré superlattices. *Nature* **597**, 350–354 (2021).
36. Lau, C. N., Bockrath, M. W., Mak, K. F. & Zhang, F. Reproducibility in the fabrication and physics of moiré materials. *Nature* **602**, 41–50 (2022).
37. Lee, J., Mak, K. F. & Shan, J. Electrical control of the valley Hall effect in bilayer MoS2 transistors. *Nat. Nanotechnol.* **11**, 421–425 (2016).
38. Wang, L. *et al.* One-Dimensional Electrical Contact to a Two-Dimensional Material. *Science* **342**, 614–617 (2013).
39. Beconcini, M., Taddei, F. & Polini, M. Nonlocal topological valley transport at large valley Hall angles. *Phys. Rev. B* **94**, 121408 (2016).

Figures

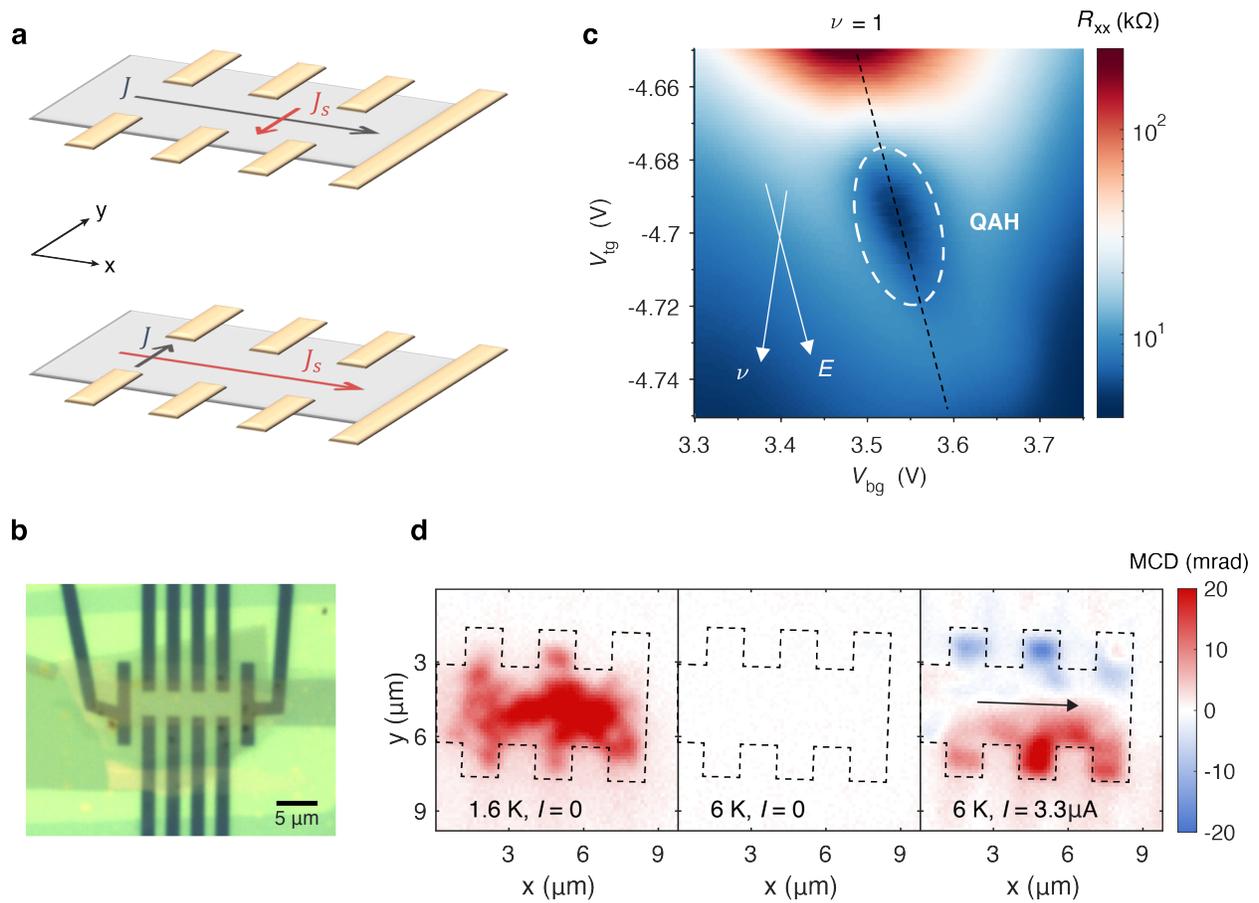

**Figure 1 | Direct observation of the spin Hall effect. a,** Schematic illustration for the local (top) and non-local (bottom) spin Hall measurements. Top, Bias current along the x-axis induces a spin Hall current along the y-axis. Bottom, Bias current between a pair of local Hall probes along the y-axis induces a spin Hall current along x-axis, enabling non-local spin Hall transport. **b,** Optical microscope image of multi-terminal Hall bar device. The scale bar represents 5 $\mu$m. **c,** Dependence of the longitudinal resistance $R_{xx}$ on the top and bottom gate voltages ($V_{tg}$ and $V_{bg}$). The arrows label the filling factor $\nu$ and electric field $E$ axes. The white dashed line encircles the QAH region at $\nu = 1$. **d,** MCD images taken at the center of QAH region ($V_{tg}$=-4.693V, $V_{bg}$=3.528V) at 1.6K with zero bias current (left panel), 6K with zero bias current (middle panel) and 6K with bias current equals 3.3$\mu$A. Black dashed lines mark the sample boundaries and arrows show the bias current direction.

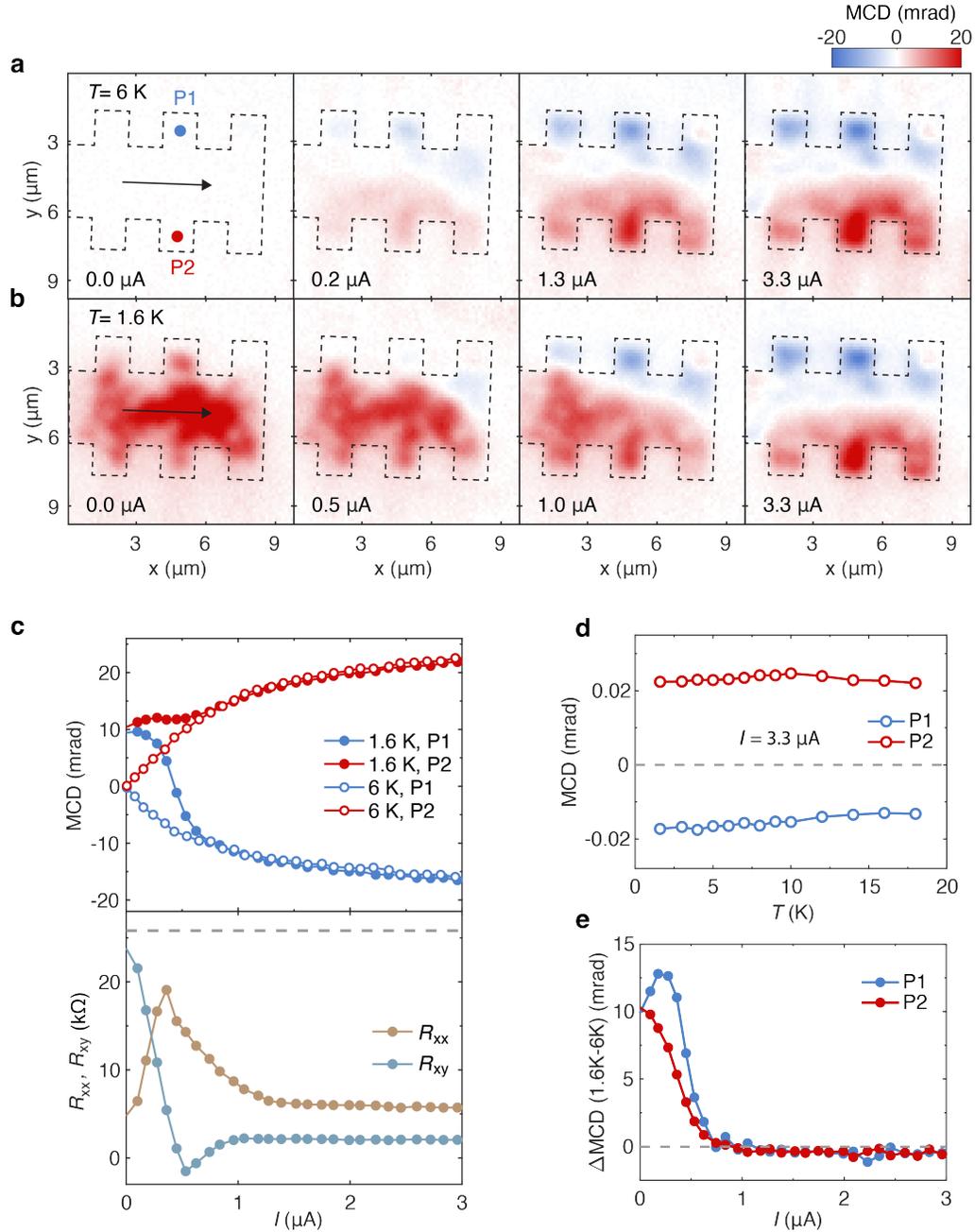

**Figure 2 | Giant spin accumulation and QAH breakdown. a,b,** Bias current- dependent MCD images at 6 K (**a**) and 1.6 K (**b**) taken at the center of the QAH region ($V_{tg}$=-4.693V, $V_{bg}$=3.528V). Black dashed lines mark the sample boundaries and arrows show the bias current direction; and P1 and P2 mark the edge locations for the data in Fig. 2. Zero-bias spontaneous MCD is observed only at 1.6 K. The high-bias MCD images, which consist of two domains, are nearly identical for 1.6 K and 6 K. **c,** Bias current dependence of edge MCD from P1 and P2 at 1.6 K and 6 K (top) and $R_{xx}$ and $R_{xy}$ at 1.6 K (bottom). A QAH breakdown is observed near 0.5 µA at 1.6 K (the horizontal dashed line marks the resistance quantum). Concurrently, the

MCD at P1 switches sign at the QAH breakdown while that at P2 increases gradually and saturates. No QAH breakdown is observed at 6 K; the MCD on both edges increase monotonically and saturates at high bias. **d,** Temperature dependence of MCD at high bias (3.3 µA) for both P1 and P2. The MCD is nearly saturated and independent of temperature at this bias current. **e,** Difference in edge MCD (ΔMCD) between 1.6 K and 6 K as a function of bias current. The ΔMCD signal reflects the spontaneous magnetization for the QAH state, which vanishes beyond the QAH breakdown.

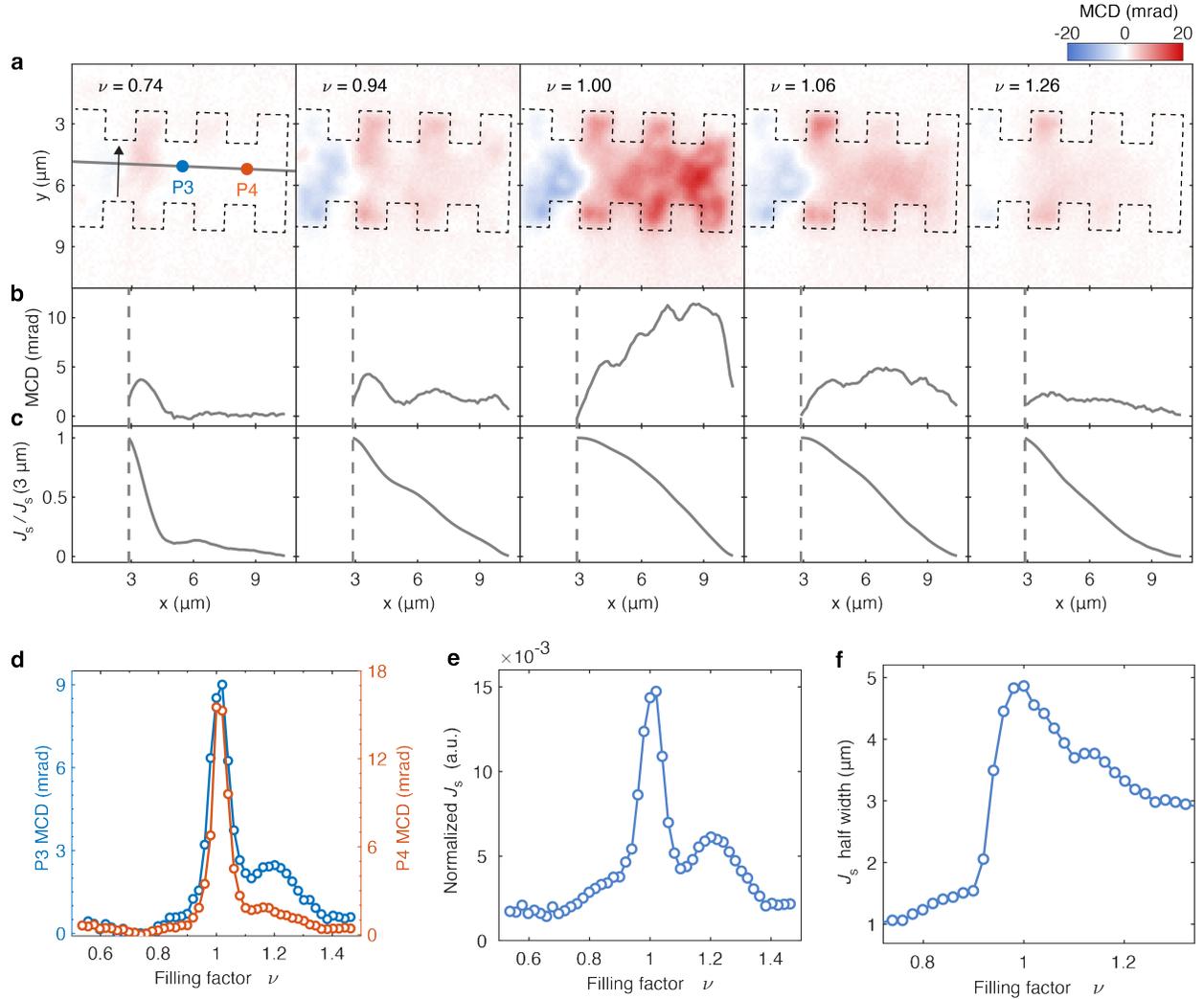

**Figure 3 | Doping dependent non-local spin Hall transport. a,** Filling factor dependent MCD images at a high bias current of 2.6 µA (beyond the QAH breakdown at 1.6 K) in the non-local measurement geometry. Black dashed lines mark the sample boundaries and the arrow shows the bias current direction. **b,c,** MCD line profile (**b**) and the corresponding spin Hall current density (**c**) at varying filling factors (taken at the grey line in **a**). $J_s$ is normalized at the current path centerline at 3µm (vertical dashed line). Non-local spin Hall transport and spin accumulation far away from the current path centerline are the most significant at $\nu = 1$. **d,** Filling factor dependence of the local and non-local MCD extracted at P3 and P4 marked in (**a**). The non-local MCD is greatly enhanced at $\nu = 1$ compared to the local MCD, where a local peak near $\nu = 1.2$ is also observed. **e,** Spin current $J_s$ at the current path centerline 3µm as a function of filling factors. $J_s$ is normalized to the resistance at each filling factor. **f,** Half-width half-maximum of the $J_s$ line profile as a function of filling factors at 1.6 K. Enhancement of the half-width is clearly observed at $\nu = 1$.

**Extended Data Figures**

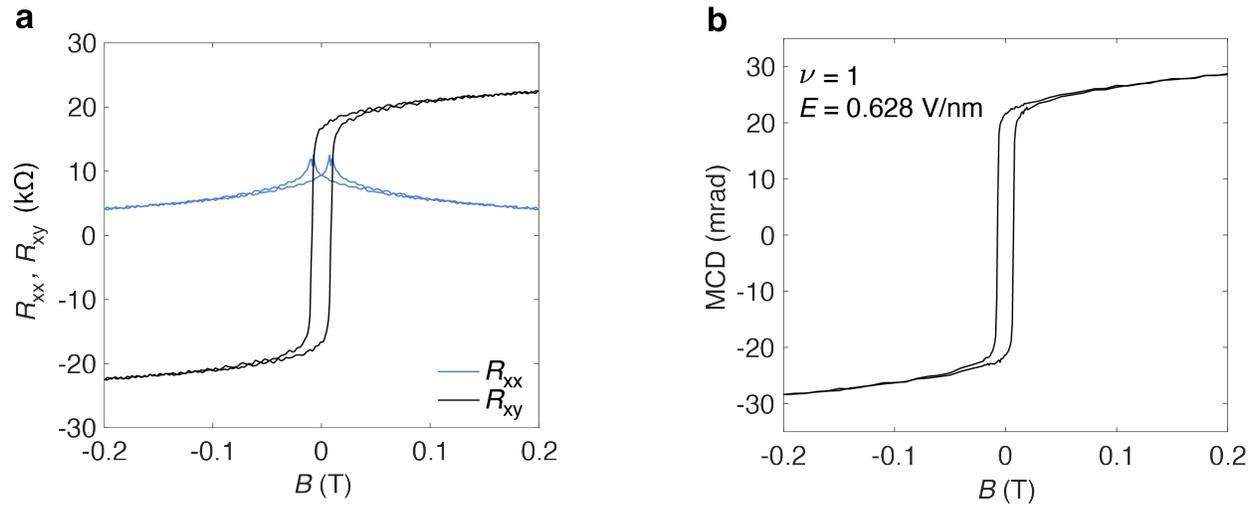

**Extended Data Figure 1 | Basic characterization for the Chern insulator. a,** Magnetic-field dependence of $R_{xx}$ and $R_{xy}$ at 1.6 K for the Chern insulating state ($V_{tg}$=-4.693V, $V_{bg}$=3.528V). Nearly quantized $R_{xy}$ and vanishing $R_{xx}$ are observed at magnetic fields higher than 0.2 T. **b,** The corresponding magnetic-field dependent MCD. A hysteresis is observed, consistent with the transport results and with the emergence of a ferromagnetic state.

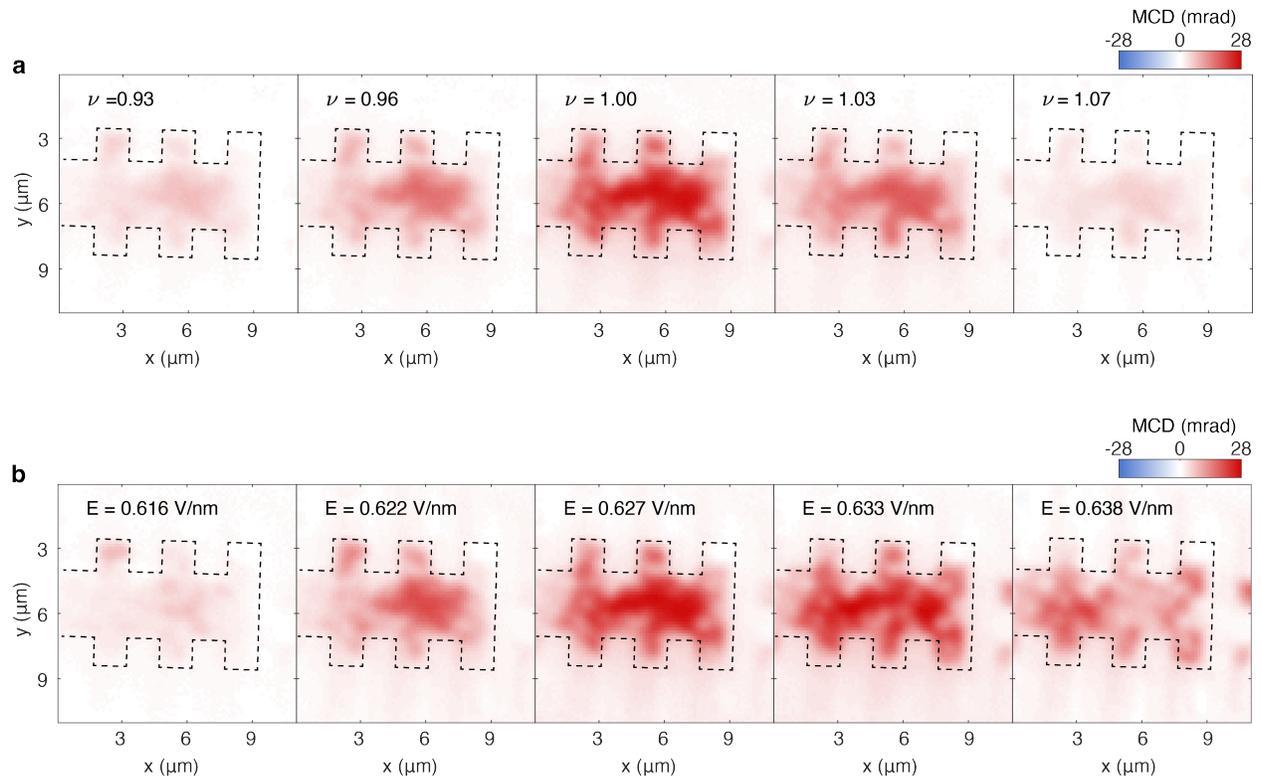

**Extended Data Figure 2 | Zero-bias MCD images at varying fillings and electric fields. a, b,** Zero-bias spontaneous MCD images at varying filling factors and fixed $E = 0.628$ V/nm (**a**) and at varying electric fields and fixed $\nu = 1$ (**b**). Temperature is at 1.6 K. Black dashed lines mark the sample boundaries. Strongest MCD is observed at the Chern insulating state ($\nu = 1$ and $E = 0.633$ V/nm); the MCD signal decreases with detuned $\nu$ and $E$. MCD inhomogeneity originating from strain and/or twist angle disorders is also observed.

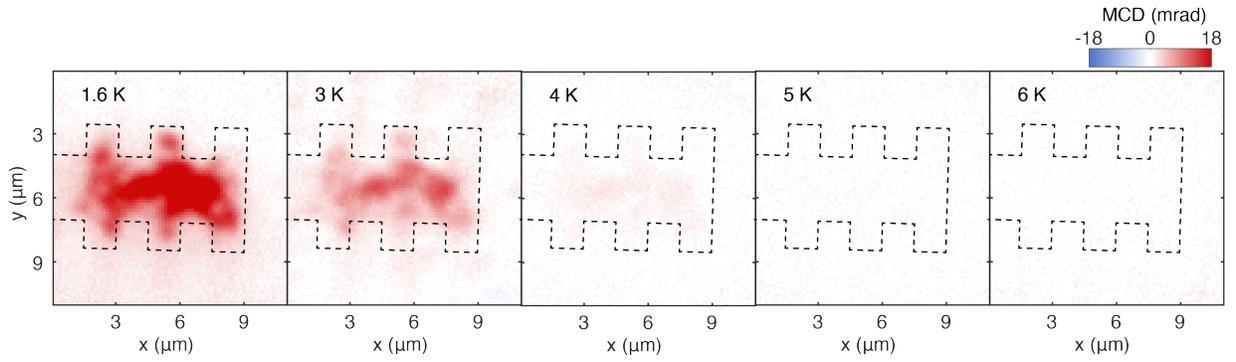

**Extended Data Figure 3 | Temperature dependent spontaneous MCD.** Zero-bias spontaneous MCD images at varying temperatures for the Chern insulating state ($V_{tg}$=-4.693V, $V_{bg}$=3.528V). Spontaneous MCD is observed below about 5 K. Black dashed lines mark the sample boundaries.

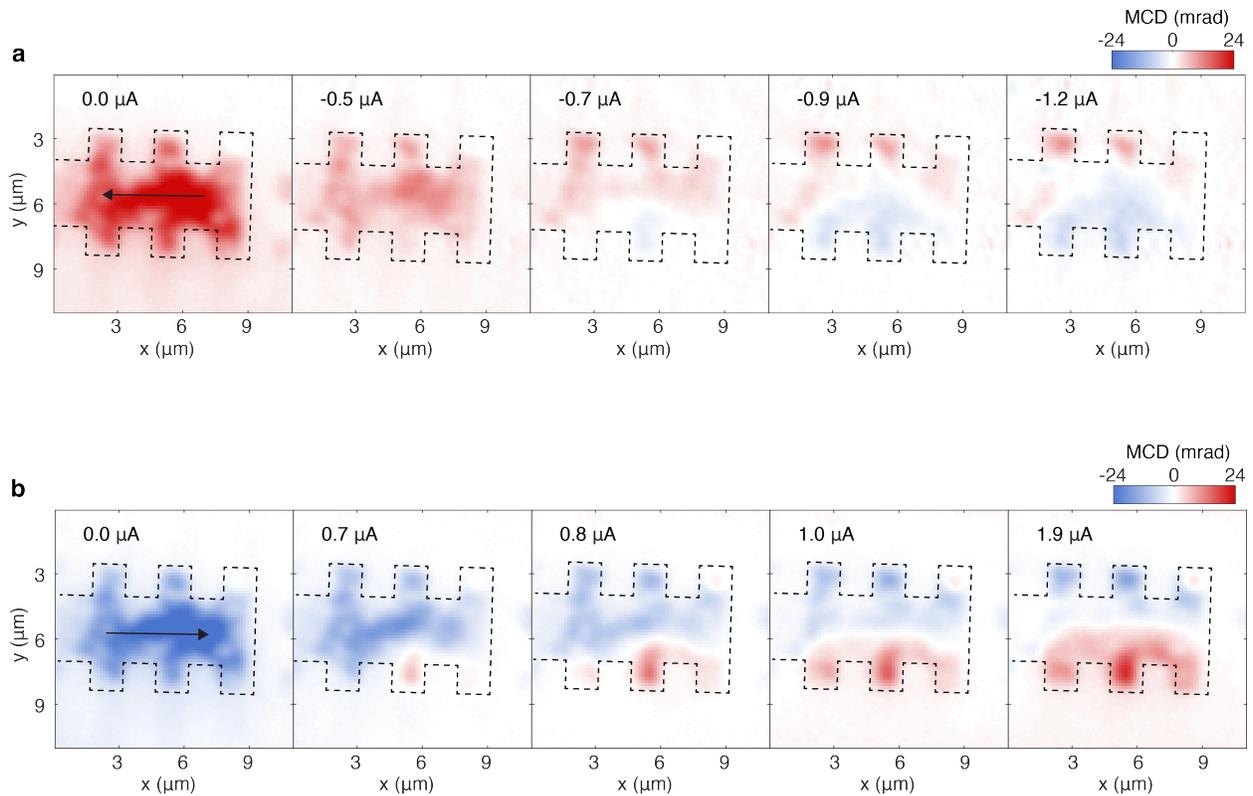

**Extended Data Figure 4 | Dependence of the SHE on the directions of bias current and spontaneous magnetization. a,** MCD images at 1.6 K under reversed bias current compared to Fig. 2b. Positive spontaneous MCD is prepared at zero bias. The current-induced edge MCD has opposite signs compared to Fig. 2b, consistent with the bulk SHE beyond QAH breakdown. **b,** MCD images at 1.6 K with spontaneous MCD opposite to that in Fig. 2b. The same current-induced MCD images are observed at high bias, demonstrating that the SHE is independent of the spontaneous magnetization direction.

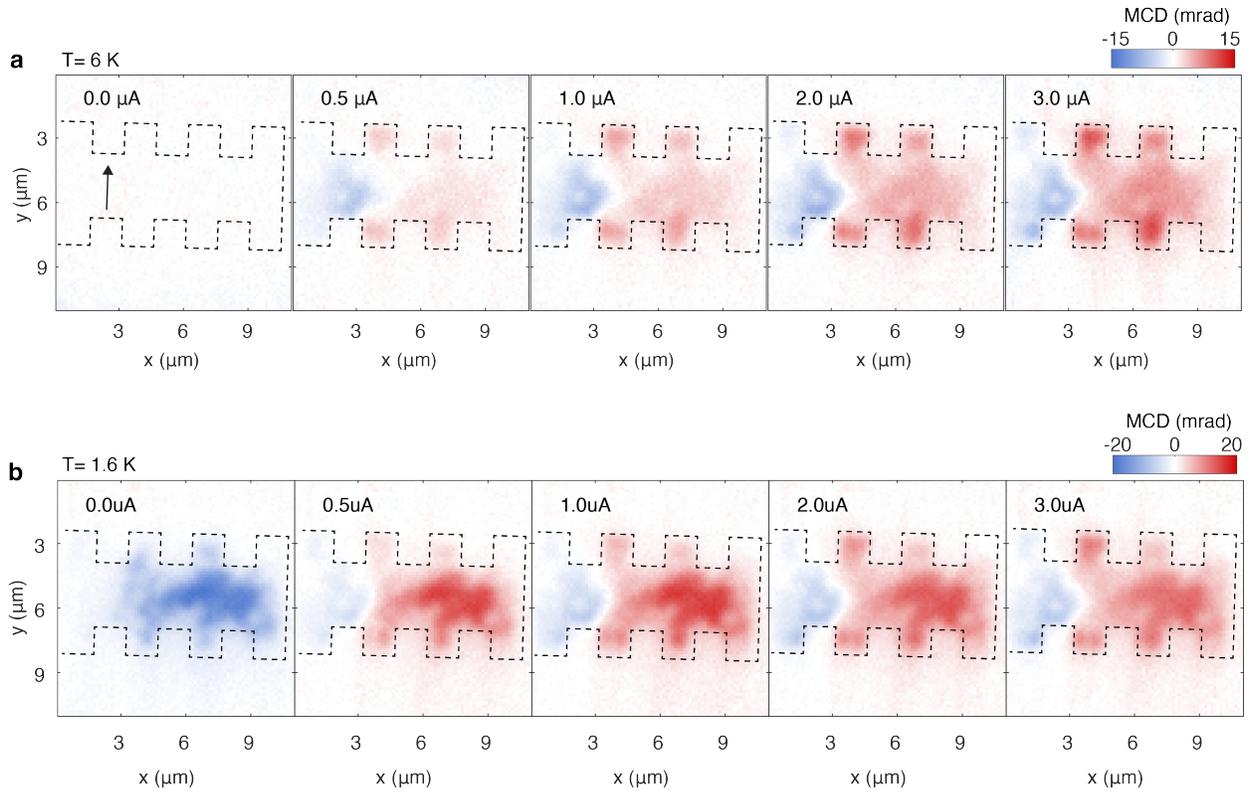

**Extended Data Figure 5 | SHE with currents biased along the short axis.** MCD images at 6 K (**a**) and 1.6 K (**b**) under varying bias currents. The bias current is along the short axis near one end of the Hall bar. The current is confined to a small region near the tip of the electrodes. Opposite magnetizations are observed on two sides of the current and the magnitude increases with increasing bias current. The current path, identified from zero MCD, deviates from a straight line due to sample inhomogeneities. Spontaneous MCD is also observed at 1.6 K under zero bias. The spontaneous MCD is quenched beyond the QAH breakdown. The 1.6 K and 6 K MCD images are nearly identical under high bias.

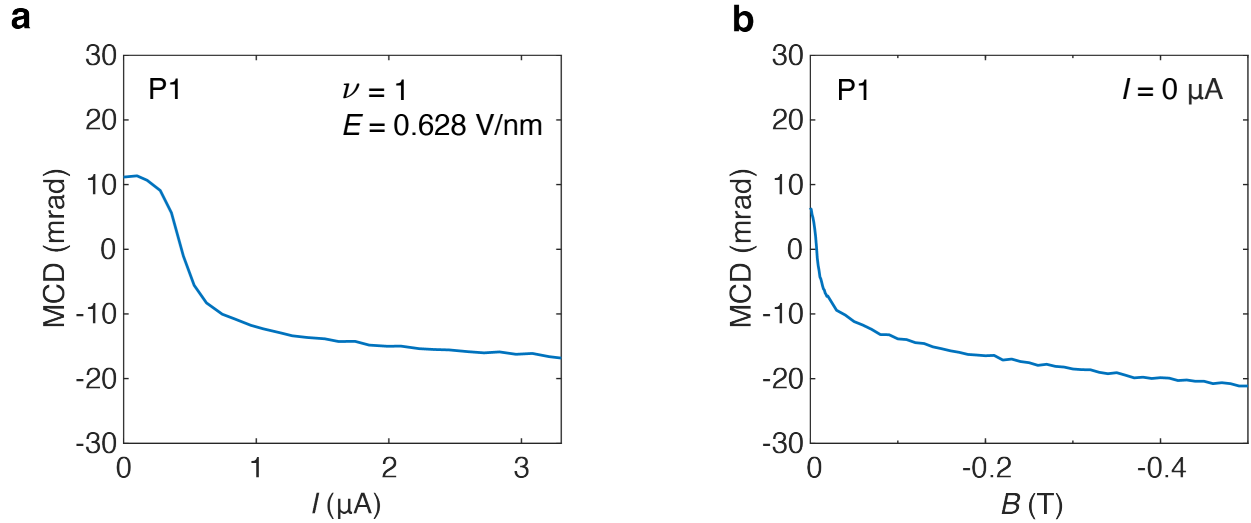

**Extended Data Figure 6 | Giant spin accumulation at high bias. a,** Bias current-dependent edge MCD at 1.6 K for the Chern insulating state (at P1 in Fig. 2a). A QAH breakdown is observed near 0.5 µA, where the MCD switches sign. **b,** The corresponding magnetic field dependent zero-bias MCD at the same sample location. We can see that the current-induced MCD at high bias (**a**) is as strong as the zero-bias MCD near magnetic saturation at 0.5 T (**b**). The results demonstrate the giant spin accumulation on sample edges due to the SHE.

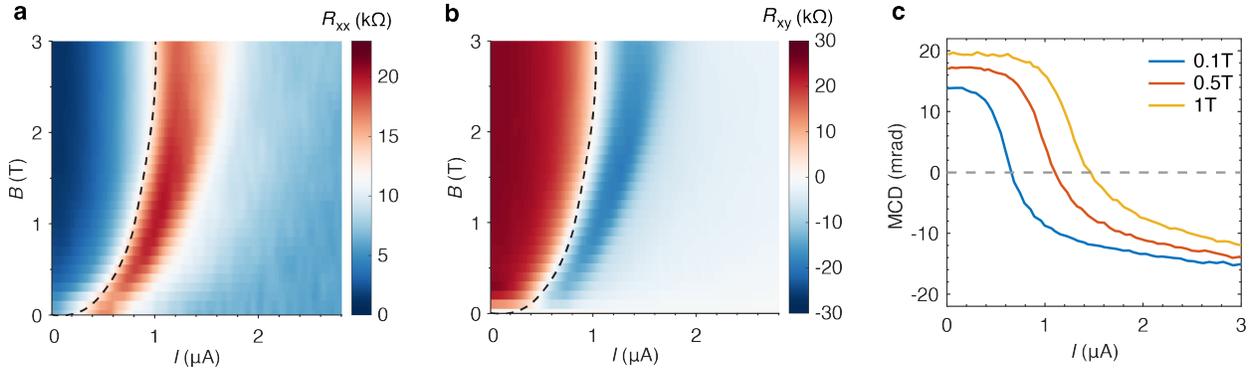

**Extended Data Figure 7 | Magnetic field dependent QAH breakdown. a, b,** Dependence of $R_{xx}$ (**a**) and $R_{xy}$ (**b**) on the bias current and magnetic field at 1.6 K for the Chern insulating state. QAH breakdown is signified by rapid changes in both $R_{xx}$ and $R_{xy}$ (the dashed lines). The critical current for the QAH breakdown increases with magnetic field. **c,** Corresponding bias current-dependent edge MCD (at P1 in Fig. 2a) at representative magnetic fields. The MCD switches sign at the QAH breakdown critical current, which increases with magnetic field, consistent with the transport results.

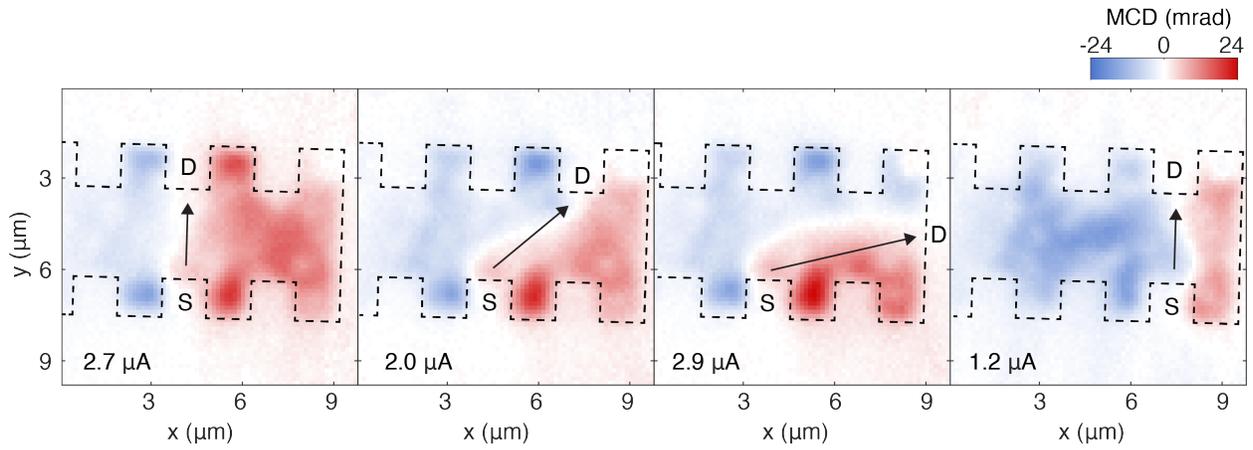

**Extended Data Figure 8 | SHE with different bias current path.** MCD images at 1.6 K and $\nu = 1$ using different pairs of Hall probes as the source (S) and drain (D). The bias current is shown in each panel. Black dashed lines mark the sample boundaries; and arrows show the bias current direction. Due to the SHE, the sample is split into two domains of opposite MCD according to the bias current path. The current path centerline, where the MCD is zero, deviates from a straight line connecting S and D due to sample inhomogeneity.

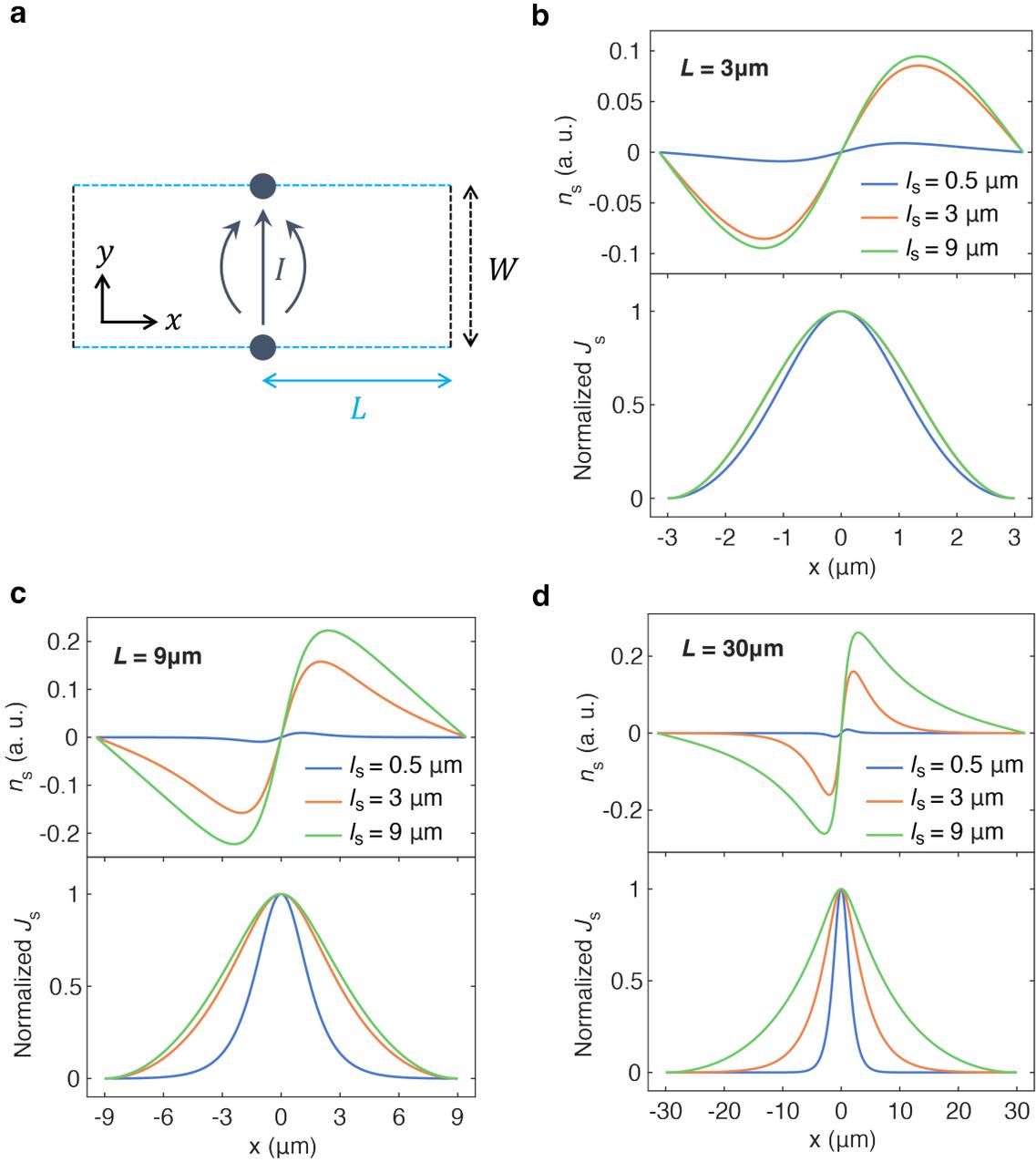

**Extended Data Figure 9 | Steady state drift-diffusion model in non-local geometry. a,** Schematic sample geometry with half-length $L$ and width $W$. The coordinate system is set up such that the sample center is at the origin. Current ejection/extraction points are marked by dark blue disks at $x = 0$, $y = \pm W/2$. **b, c, d,** Spin density $n_s$ and spin current density $J_s$ profile at $y = 0$ for representative spin diffusion lengths with sample geometry $L = 3\mu m$ (**b**), $9\mu m$ (**c**), $30\mu m$ (**d**) and fixed $W = 3\mu m$. A Hall angle of $\pi/6$ is used in all cases. With increasing $L$, the spin density becomes more localized around the current centerline and the spin current density $J_s$ shows stronger dependence on the spin diffusion length.